\documentclass[12pt]{iopart}
\usepackage{graphicx}
\usepackage{bm}      
\usepackage{epsfig}
\usepackage{wasysym}
\usepackage{cases}
\usepackage{fullpage}
\usepackage{setspace}

\begin{document}

\title{A disorder induced field effect transistor in bilayer and trilayer graphene}

\author{Dongwei Xu$^{1}$,  Haiwen Liu$^{2,3}$, Vincent Sacksteder IV$^{2}$, Juntao Song$^{4}$, Hua Jiang$^{3,5, \ast}$, Qing-feng Sun$^{2}$ and  X. C. Xie$^{3}$}

\address{$^1$Department of Physics, Oklahoma State University, Stillwater, Oklahoma 74078, USA;\\
$^2$Institute of Physics, Chinese Academy of Sciences, Beijing 100190, China;\\
$^3$International Center for Quantum Materials, Peking University, Beijing 100871, China;\\
$^4$Department of Physics and Hebei Advanced Thin Film  Laboratory, Hebei Normal University, Hebei 050024, China;\\
$^5$Department of Physics, Soochow University, Suzhou, 215006, China}

\ead{jianghuaphy@suda.edu.cn}

\begin{abstract}
We propose use of  disorder  to produce a  field effect transistor (FET) in biased bilayer and trilayer graphene.  Modulation of the bias voltage can produce large variations in the conductance when the disorder's effects are confined to only one of the graphene layers.   This effect is based on the bias voltage's ability to  select which of the graphene layers carries current, and is not tied to the presence of a gap in the density of states.  In particular, we demonstrate this effect  in models of gapless ABA-stacked trilayer graphene, gapped ABC-stacked trilayer graphene, and gapped bilayer graphene.
\end{abstract}


\maketitle

\section{Introduction}

When a bias voltage is applied perpendicularly  to a sheet of bilayer
graphene or rhombohedrally (ABC) stacked trilayer gaphene, it opens an
energy gap and regulates the gap's size ~\cite{ohta2006science,
abergel2010AP,taychatanapat2010PRL, guinea2006PRB, zhang2010PRB,
avetisyan2009PRB, jhang2011PRB}.  This remarkable effect  allows the
realization of digital logic based on graphene-based switchable
devices such as field effect transistors (FETs).   Turning on the
graphene FET's bias voltage creates a band gap which in turn halts
electronic conduction and switches the FET from the on (conducting)
state to the off (insulating) state.    Such devices first were
studied theoretically ~\cite{mccann2006PRB,min2007PRB,yu2008PRB} and
later were demonstrated experimentally ~\cite{castro2007PRL,
zhang2009Nature, weitz2010Science,
oostinga2008NM,Miyazaki10,Yamashiro12, Shioya12}.   Xia \textit{et al}
fabricated a dual-gated FET with a transport gap of $ > 130$ meV,
which compares favorably to room temperature ($26$ meV)
~\cite{xia2010NL} . They achieved an on/off current ratio of  $100$ at
room temperature and $2000$ at $20$K.  More recent experiments used
doping to create a mobility gap in  single-gated bilayers
\cite{Szafranek11},   adjusted the Fermi level via doping \cite{Yu11},
 and demonstrated a $235$ meV transport gap via ionic gating
\cite{Yamashiro12}. San-Jose \cite{San-Jose09PRL}  proposed a concept of pseudospin valve in bilayer graphene based on the electron transport controlled by the external gate.  Some initial studies of transport in dual gated
graphene trilayers  also have been reported ~\cite{jhang2011PRB,craciun2009NN}. The principal motivation for these studies is the
promise of graphene-based digital devices.

\begin{figure}
\begin{center}
 \includegraphics[width=\columnwidth, viewport=15 12 630 155, clip]{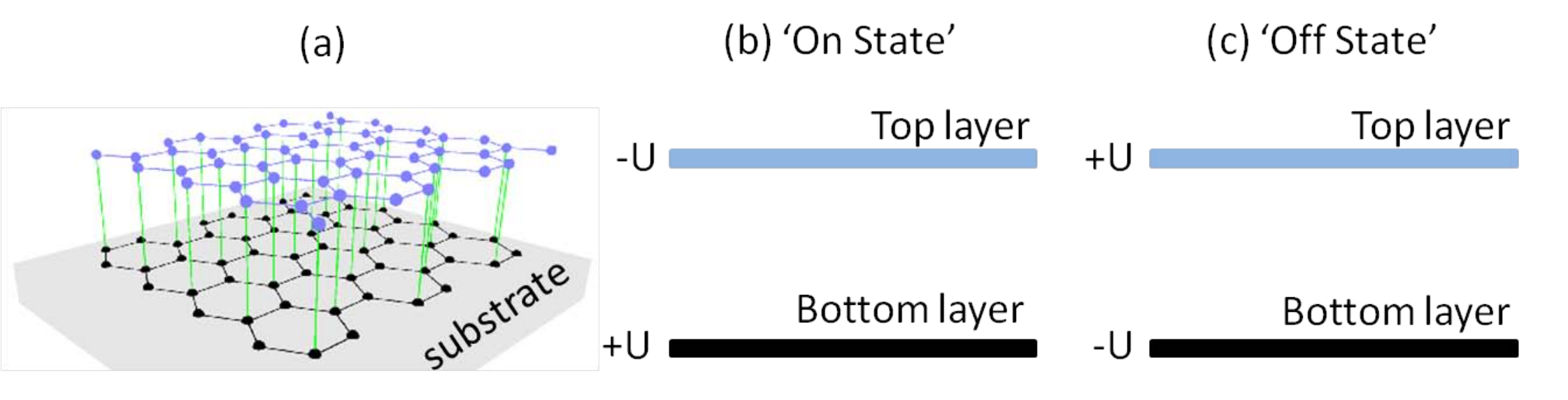}
  \end{center}
  \caption{(Color online.) Schematic diagram of the proposed graphene-based transistor.  Pane (a) shows  the graphene bilayer and its supporting disordered substrate.   The black coloring represents disorder which strongly affects the lower layer and only weakly affects the upper layer.  A bias voltage is applied perpendicularly to the bilayer. In pane (b) the bias voltage raises the  energy of the lower layer by $+U$ and lowers the voltage of the upper layer by $-U$. In pane (c) the bias voltage is reversed. }
  \label{fig:schematic_diagram}
\end{figure}

In this paper we  propose an alternate architecture for graphene transistors which does  not require  a band gap.   Figure ~\ref{fig:schematic_diagram}a illustrates the  graphene devices which we propose:  graphene bilayers (or trilayers) supported by a disordered substrate,  with strong disorder in the lower graphene layer and much weaker disorder in the upper layers.  Current flows longitudinally through the graphene, and a perpendicular  bias voltage switches between  the 'on' (conducting) state and the 'off' state.

The key physics of our FET architecture lies in the density of states, whose value at  the Fermi level $E = E_F$ determines the carrier density and the conductance. We will show that in  bilayer and trilayer graphene a bias voltage can cause the electrons at the Fermi level to concentrate on one or the other of the graphene layers, so that conduction favors one layer over the other.    If the conduction electrons concentrate on the strongly disordered lower graphene layer, the conductance will be strongly attenuated by the disorder.     Reversing the bias voltage will shift the  density of states  to concentrate on the upper graphene layer which is only weakly affected by disorder and therefore has a much larger conductance.

In this paper we  will focus on computing and understanding the  conductance ratio $G_{+U}/G_{-U}$, where the longitudinal conductances $G_{\pm U}$ are measured between two edges of the graphene sheet.  The numerator in this ratio - the positive-biased conductance -  is illustrated in   figure ~\ref{fig:schematic_diagram}b where a positive bias voltage elevates the energy of the lower layer.  The denominator is shown in figure ~\ref{fig:schematic_diagram}c where the bias voltage is reversed and depresses the lower layer's energy.    A large value of $G_{+U}/G_{-U}$ indicates good switching between on and off states. The main contributions of our paper are that (1) disorder which is concentrated on the lower graphene layer can dramatically increase the conductance ratio, and (2) a large conductance ratio can be obtained even when  there is no bulk band gap.

Our proposed  disorder-based FET converts disorder from  a liability to a resource and allows greater flexibility in FET design.  In particular, it allows use of multilayers which do not develop a band gap when a bias voltage is applied.  Trilayer graphene has two stable allotropes which are distinguished by their stackings.  The more naturally abundant allotrope follows  the Bernal (ABA) stacking order which possesses both inversion and mirror symmetry, and these symmetries prevent creation of a band gap.  Previous work on gap-based trilayer graphene FETs has avoided the  ABA stacked allotrope and concentrated on the less abundant rhombohedrally (ABC) stacked allotrope which permits creation of a band gap ~\cite{guinea2006PRB, kumar2011APL,lui2011NP,bao2011NP,yacoby2011NP,koshino2009PRB, min2008PTP} . We will show that a disorder-based FET can be realized with both trilayer allotropes.

Furthermore, our disorder-based FET gives new possibilities for engineering  the width of the 'off' state and the Fermi level where it occurs.  In gap-based FETs the Fermi level must be tuned to coincide with the gap, which occurs at graphene's Dirac point and has a width that is controlled by the  multilayer's band structure.  A small gap or improper tuning of the Fermi level would restrict the working temperature to small values and decrease the conductance ratio.  In contrast, in our architecture the 'off' Fermi level is controlled by the bias voltage;  the bias voltage can be changed to tune the 'off' state to coincide with the Fermi level.  We  will show that properly engineered disorder-based FETs can have 'off' state widths of order $100-200$ meV, which is competitive with  gap-based FETs.   In effect,  compared to gap-based FETs which are completely determined by band structure,  disorder-based FETs are much more  flexible and controllable.  They are sensitive to  design of a suitable substrate and of its bonding with graphene,  to engineering of the disorder type and strength, and to choice of Fermi level and bias voltage.

Our FET proposal occurs in the context of an already rich theoretical literature about conduction in disordered bilayer and multilayer graphene \cite{Nilsson08,Adam08,Koshino09,Ferreira11,Yuan10,Zare10,Xu11,Xu09,Castro10,Min11,Rossi11,Wakatsu12,Trushin10,DasSarma11,Koshino08,Culcer09,McCann12,Kechedzhi12}.  Various previous works examined many different types of disorder, calculated the density of states and the scattering length, included screening effects, and paid careful attention to the conductance's dependence on the electron density at both high and low densities.  In general the focus has been on very small defect densities, because disorder has been understood as detrimental to device performance and experimentalists have had considerable success in avoiding defects and increasing the mobility.

In contrast to gap-based graphene  FETs,  our FET architecture requires disorder to  affect the lower graphene layer more than the upper layer, breaking their symmetry.  This disorder profile is physically plausible if the lower layer is supported by a substrate or is subjected to a disordering treatment, but has been studied very little \cite{Zhong07PRB, Mao10NJP, Xu08}. Zhong and collaborators \cite{Zhong07PRB, Mao10NJP} showed that if disorder is confined to one layer then some portions of the spectrum remain in the conducting  phase  even at very large disorder.   Mobility edges separate the insulating and conducting phases. Here we explore the possibility of using this disorder profile in conjunction with a gate induced bias voltage to realize an FET.

We have studied FET performance as the disorder strength is varied by an order of magnitude.  Although we found some weak FET-like behavior at smaller disorder strengths, we found acceptable conductance ratios only when  the 'off' state  lies near the Anderson localization regime; strong disorder is required.   Each scatterer must be quite strong, placing it near  the unitary scattering limit, similarly to vacancies and resonant scatterers.  A variety of strong nonlinear effects including charge redistribution between layers may occur in  this strong disorder regime.  In the lower graphene layer there must be  a high density of scatterers, which again contrasts with previous studies.

     Because our target disorder regime has not been thoroughly studied in multilayer graphene, and because there is   ample room for tailoring the disorder and the substrate to suit the FET,  in this paper we have adopted a  simplified - even phenomenological - approach which is appropriate for exploring this architecture's possibilities. We have not made a detailed analysis of the many possible types of disorder, examined  charge equilibration, incorporated the details of the gate potential or the leads, or the like. The present paper is a proof of concept which focuses on  two key mechanisms which can cooperate to produce an FET;  we show (A) that gating can shift the carriers to one layer or another, and (B) that the conductance is reduced when the carriers are shifted to the more disordered graphene layer.    Our proof of (A) is quite simple, based on $\vec{k} \cdot \vec{p}$ model calculations of pristine graphene's band structure.    These results are supported by our tight-binding calculations which incorporate disorder.  Our proof of (B) is based on the same  tight-binding calculations.   Our tight binding model confirms that large conductance ratios can be achieved in bilayers and trilayers with strongly disordered lower  layers.

  \Sref{sec:model}  describes our tight binding models and numerical method. \Sref{sec:RD}.1 uses the tight binding model of bilayer graphene to calculate the  conductance ratio, and analyzes the effect of disorder strength, bias voltage, two disorder types, and edge type on the FET's effectiveness.    \Sref{sec:RD}.2 explores the band structure of bilayer graphene and explains the conductance ratio in terms of  a shift in the local electron distribution between the upper layer and the disordered lower layer.  In \sref{sec:RD}.3 we  investigate the conductance ratio and band structure of both ABA-stacked and ABC-stacked trilayer graphene, and show that our proposed FET does not require the existence of a band gap.  \Sref{sec:conlusion} summarizes our conclusions and presents opportunities for further research.

\section{Tight binding calculation of the conductance}\label{sec:model}

We model bilayer and trilayer graphene with Hamiltonians that include explicitly the electronic states in the graphene layers.   The FET architecture is shown in  \fref{fig:schematic_diagram}a, and the 'on' and 'off' states are indicated   in \fref{fig:schematic_diagram}b and  \fref{fig:schematic_diagram}$c$.  We use well known tight-binding Hamiltonians ~\cite{nilsson2008PRB, qiao2011PRL, neto2009RMP}  for Bernal (AB) stacked bilayer graphene and for  ABA  (Bernal) and ABC (rhombohedral) stacked trilayers:
\begin{eqnarray}
H_{AB} &=&\sum\limits_{\langle i,j\rangle ;L=B,T} -t_{ij,L}\left( a_{i,L}^{\dag
}b_{j,L}+h.c.\right)
+\sum\limits_{i;L=B,T;c=a,b}U_{i,L}c_{i,L}^{\dag }c_{i,L}
\nonumber \\
&& - t_{\bot }\sum\limits_{i}\left( b_{i,T}^{\dag }a_{i,B}+h.c.\right)
\end{eqnarray}%
\begin{eqnarray} \label{ABATightBinding}
H_{ABA} &=&\sum\limits_{\langle i,j\rangle ;L=B,M,T} -t_{ij,L}\left( a_{i,L}^{\dag
}b_{j,L}+h.c.\right)
+ \sum\limits_{i;L=B,T;c=a,b}U_{i,L}c_{i,L}^{\dag }c_{i,L}
\nonumber \\
& &- t_{\bot }\sum\limits_{i}\left( b_{i,T}^{\dag }a_{i,M}+a_{i,M}^{\dag
}b_{i,B}+h.c.\right)
\end{eqnarray}
\begin{eqnarray} \label{ABCTightBinding}
H_{ABC} &=&\sum\limits_{\langle i,j\rangle ;L=B,M,T} t_{ij,L} \left( a_{i,L}^{\dag
}b_{j,L}+h.c.\right)
+ \sum\limits_{i;L=B,T;c=a,b}U_{i,L}c_{i,L}^{\dag }c_{i,L}
\nonumber \\
& &-t_{\bot }\sum\limits_{i}\left( b_{i,T}^{\dag }a_{i,M}+b_{i,M}^{\dag
}a_{i,B}+h.c.\right)
\end{eqnarray}
In each Hamiltonian the first term describes intralayer hopping, the second term describes the bias potential, and the third term describes interlayer hopping.  The $L$ index selects the graphene layer, and has values `B', `M', and `T' for the lower, middle and upper layers respectively.  Each graphene layer is composed of two sublattices;  $a_{i}$ and $b_{i}$ are the annihilation operators of the two sublattices.  Intralayer hopping occurs only between nearest neighbors, as signaled by the $\sum\limits_{\langle i,j\rangle}$ notation.

We choose the parameters of pristine graphene as follows: $t$, the intralayer hopping strength, is a constant $t = 2.7$ eV.  Throughout this paper all energies will be quoted in units of $t$.  $t_{\bot}$, the interlayer hopping strength, is a constant $t_{\bot} = 0.1496 t$.  The bias voltage $U$ determines the on-site potential $U_{i,L}$.  When calculating the 'on' conductance $G_{+U}$ we set $U_{i,B}$ on the lower layer to $+U$ and $U_{i,T}$ on the upper layer to $-U$.  When calculating the 'off' conductance the signs are reversed.  The on-site potential of trilayer graphene's middle layer is always zero.

We model the two leads as being semi-infinite and connected at opposite ends of the FET.  Their structure, Fermi level, and bias voltage  are identical to those of the FET itself; the only difference is that leads are clean while the FET is disordered.  There is no lattice mismatch between the leads and the FET.

We evaluate the longitudinal conductance using the nonequilibrium Green's function formalism ~\cite{Meir92,datta1995, long2008PRL, jiang2009PRL}.  For our calculation of the conductance  without interactions this formalism is equivalent to the Caroli formula \cite{Meir92, Caroli71} $G = -\frac{e^2}{h}{Tr}((\Sigma^r_{L}-\Sigma^a_{L}) G^r_{LR} (\Sigma^r_{R}-\Sigma^a_{R}) G^a_{RL})$, where $G^a, G^r$ are the advanced and retarded Green's functions connecting the left and right leads and $\Sigma_{L,R}$ are the self-energies of the leads.  The Green's functions and the lead self-energies are evaluated recursively \cite{Sancho85}, decreasing the computational burden.

All of our calculations set the FET width and length at about  $149a$ and $344a$, where $a=0.142$ nm is the distance between nearest neighbors in graphene.

\subsection{Modeling the Disorder}
We add non-magnetic disorder to the tight binding Hamiltonians which we have just discussed.     A fundamental requirement of our FET architecture is that the disorder must break the symmetry between the lower layer and the upper layers.    In our simplified model the disorder affects only the lower graphene layer.

We consider two types of disorder.  The first is on-site Anderson disorder, a random local potential acting on individual atoms; we add random numbers to the lower layer's on-site potential $U_{i,T}$.  The second type is intralayer bond disorder, which changes the bonds within the lower graphene layer; we add random numbers to the lower layer's intralayer hopping strength $t_{ij,L}$.  As is conventional, the disorder is  randomly distributed in the interval $\left[ -W/2,W/2\right] $, where $W$ is the disorder strength.  It  is also short-range, with no correlation between neighboring sites.   All of our conductance calculations are averaged over two hundred random disorder realizations for each value of the disorder strength, Fermi level, and bias voltage.  The sample to sample variation in the conductance is  never larger than one conductance quantum $e^2 / h$.

 For small disorder strengths the bond disorder can be understood as modeling variations in the lower graphene layer's covalent carbon bonds, which may be caused by a lattice mismatch between the graphene and its substrate ~\cite{fuchs2007PRL}, or by variations in the chemical bonding between the two  ~\cite{kang2008PRB, zhou2007NM, giovannetti2007PRB}.  We will find the best FET performance at large disorder strengths $W$ which are of the same order of magnitude as the intralayer hopping $t$.    These very large variations in the bond strength demonstrate the disruption of the carbon bonds by large densities of carbon vacancies, resonant scatterers, missing bonds, adsorbed molecules,  etc.    At large disorder strength both  on-site Anderson disorder and  bond disorder are close to the unitary limit of strong short-range scatterers.  They belong to the same universality class, and we will find that they have the same effect on the conductance.

\section{Numerical results}\label{sec:RD}
In this section we will calculate the conductance ratio in bilayer and trilayer graphene.  We will examine the effects of  disorder strength and bias voltage.  We begin with bilayer graphene with zigzag edges.

\subsection{Electron transport in bilayer graphene}
\begin{figure}
\begin{center}
 \includegraphics[width=\columnwidth,viewport=15 450 2257 1700, clip]{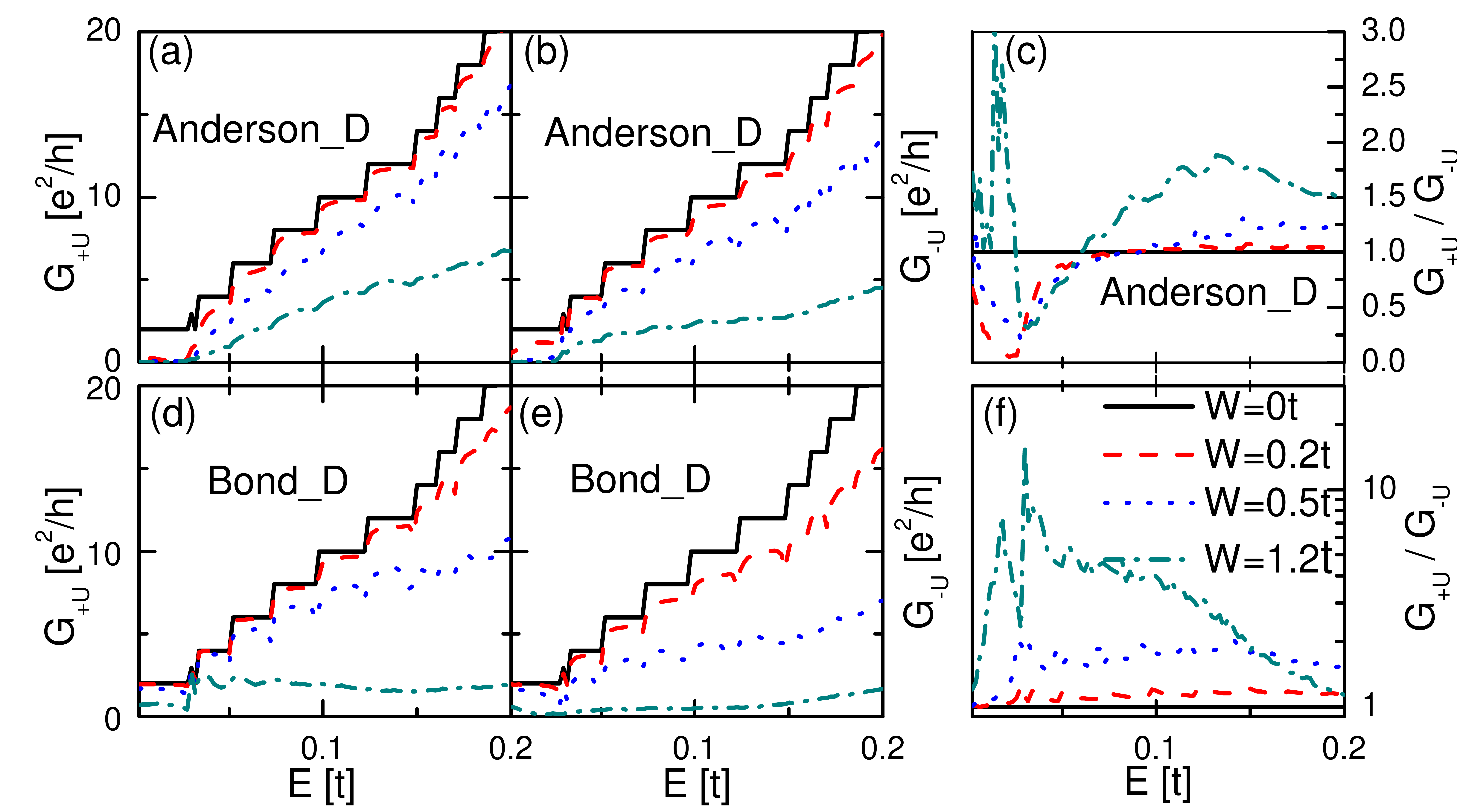}
   \end{center}
   \caption{
(Color online.)  Bilayer graphene with zigzag edges under a small  bias potential $U = \pm 0.03 t$.  Panes (a) and (d) show the conductance under a positive bias potential $U= +0.03 t$ (see  \fref{fig:schematic_diagram}b), while panes (b) and (e) show the conductance under an opposite negative bias potential $U = -0.03 t$ (\fref{fig:schematic_diagram}c).  Panes (c) and (f) show the ratio of the two conductances, $G_{+U}/G_{-U}$.  The upper panes (a), (b), and (c) were computed using on-site Anderson disorder, while the lower panes (d), (e), and (f) were computed using bond disorder.  The disorder strengths are reported in pane (f), and $t$ is the intralayer hopping matrix element.  The $x$ axis in all panes is the Fermi level.    \label{fig:example}}
\end{figure}

 \Fref{fig:example}  illustrates the conductance and conductance ratio of bilayer graphene with zigzag edges under a small bias voltage $U= \pm 0.03t$.  The left-most panes (a) and (d) show the conductance under a positive bias voltage (see  \fref{fig:schematic_diagram}b) and  the middle panes (b) and (e) show the conductance under a negative bias voltage (see  \fref{fig:schematic_diagram}c). The right-most panes (c) and (f) show the ratio $G_{+U}/G_{-U}$ of the two conductances, which governs the FET's effectiveness.  The upper panes used on-site Anderson disorder, while the lower panes used bond disorder.  The $x$ axis gives the Fermi level.

In each pane the four lines represent four different disorder strengths.  At zero disorder the conductance is a staircase between plateaus of quantized conductance. Disorder smooths the plateaus,   progressively reduces the conductance across the whole spectrum, and for on-site disorder quickly reduces the conductance at $E_F = 0$ to zero.   For large disorder $W = 1.2 t$ the conductance is reduced close to the localized regime, which is characterized by $G \leq 1$.  The conductance in the lower panes is smaller than in the upper panes, implying that bond disorder has a stronger effect that on-site Anderson disorder.

Comparing the left  panes (a) and (d) with positive bias voltage $U = +0.03 t$ to the middle panes (d) and (e) with negative bias voltage $U = -0.03 t$ quickly reveals the bias voltage's most notable affect: a negative bias voltage magnifies the disorder and reduces the conductance.  In the next section we will demonstrate  that  a negative bias can shift the density of states into the disordered lower layer, thus reducing the conductance.   \Fref{fig:example}'s right panes (c) and (f) show the  ratio between the $U = +0.03 t$ conductance and the $U = - 0.03 t$ conductance.  In general the conductance ratio is larger than one, confirming that a negative bias voltage does strengthen the disorder's effects and reduce the conductance. (The exceptional region in pane (c) is connected to small disorder strengths and small conductances.)   The conductance ratio increases when the disorder strength is increased, and reaches peaks as high as $3.0$ for on-site disorder and $10$ for bond disorder.   A trend towards $1$ is visible for large Fermi levels $E_F > 0.15$.  At large disorder the conductance ratio peaks are located near $E_F = U = 0.03 t$.   If the Fermi level lies near these peaks, then inverting the bias voltage will cause a substantial decrease in the conductance.

These results at small bias voltage $U = \pm 0.03 t$ show that bilayer graphene's conductance is sensitive to a bias voltage. However   conductance ratios higher than three are achieved only with strong $W = 1.2 t$ bond disorder.  In this case panes (d) and (e) indicate that both  the 'on' state and the 'off' state lie near the localized regime, and that the 'on'  conductance  is small $G_{+U} \propto 2$.  This scenario is not suitable for producing standardized and reproducible FETs, because the sample-to-sample  variation in the conductance is generally about one conductance unit.  'on' conductances exceeding ten conductance units are desirable.  Moreover we would like to heighten and broaden the peak in the conductance ratio. Our next results will show that both  of these requirements can be met by increasing the bias voltage.

\begin{figure}
\begin{center}
 \includegraphics[width=\columnwidth,viewport=56 37 700 534, clip]{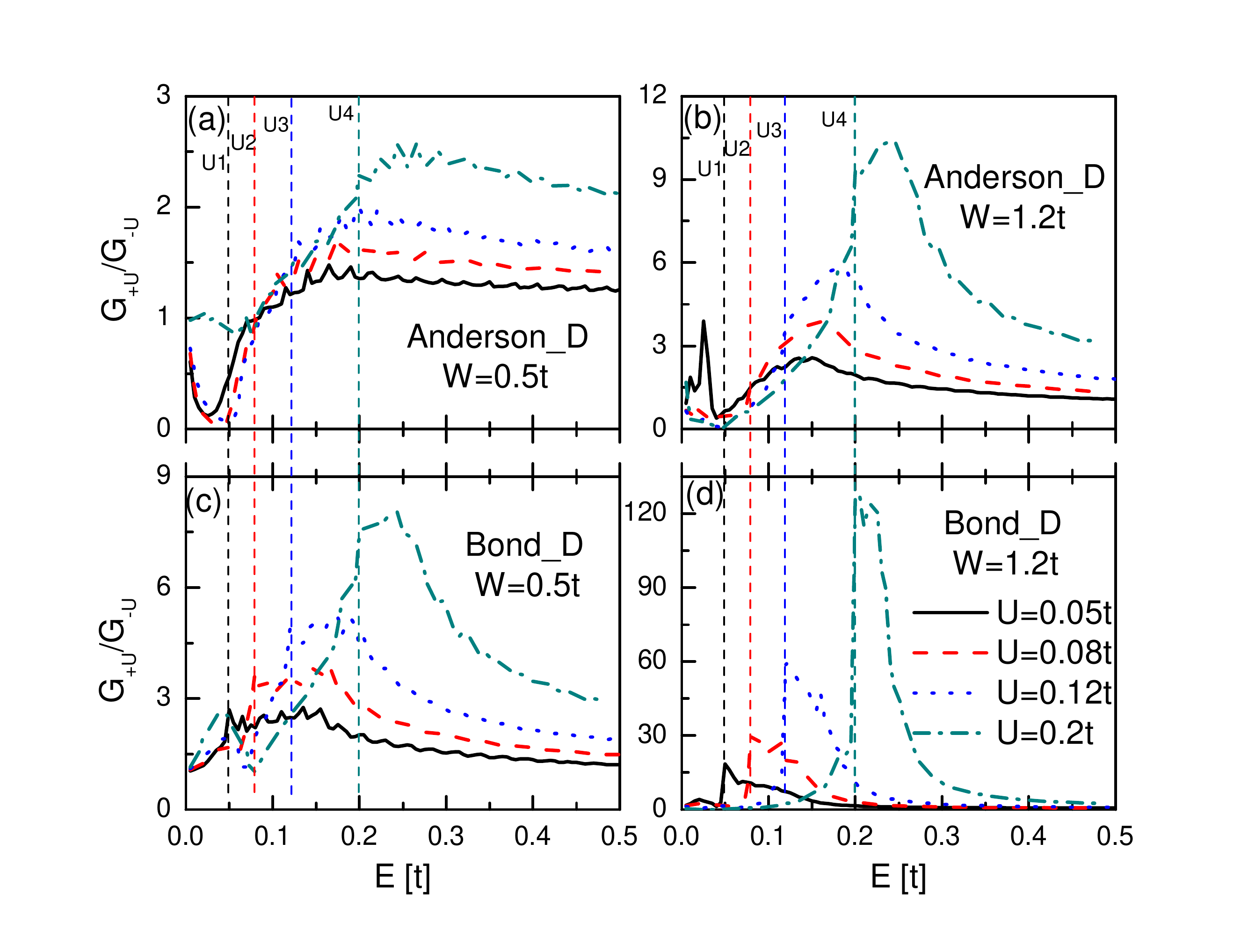}
   \end{center}
   \caption{
 (Color online) Conductance ratio $G_{+U}/G_{-U}$ of bilayer graphene with zigzag edges, under four different bias voltages.   The highest curves were obtained with bias voltage $U = \pm 0.2 t$, while the lowest curves were obtained with $U = \pm 0.05 t$. Strong disorder is shown in the right panes (b) and (d), while intermediate disorder is shown in the left panes (a) and (c).  The upper panes were obtained with on-site Anderson disorder, while the lower panes were obtained with bond disorder.  The $x$ axis shows the Fermi level.  The vertical lines labeled
$U1$, $U2$, $U3$ and $U4$ mark the point where the Fermi level is equal to the bias voltage $E_F = U$.  This point controls the left edge of the peak in the conductance ratio.
   }
  \label{fig:bilayer}
\end{figure}

\Fref{fig:bilayer} shows the  bias voltage's effect on the conductance ratio at fixed disorder strength.  The four curves in each pane correspond to four different values of the bias voltage.    The lowest curve is obtained from $U = \pm 0.05 t$, while the highest curve is obtained with $U = \pm 0.2 t$.  Again we see that positive bias voltages generally produce higher conductances than negative biases do, with the only exceptions lying at small Fermi level and small disorder.    Moreover we see as before that the conductance ratio tends toward one at large values of the Fermi level.  The upper panes show results obtained with on-site Anderson disorder, while the lower panes were obtained with bond disorder.  In the left panes we used disorder strength $W = 0.5 t$, while in the right panes we used $W = 1.2 t$.  Pane  (b) with $W = 1.2 t$ on-site disorder is a close match to pane (c) with $W = 0.5$ bond disorder, which indicates that bond disorder with strength $w$ is roughly equivalent to on-site disorder with strength $2 w$.  This equivalence is natural since both disorder types belong to the same universality class, and since these strong disorder values are close to the limit of unitary scattering.  In subsequent numerical results we will use only bond disorder.

Examination of \fref{fig:bilayer} shows that increasing the bias voltage heightens the peak in the conductance ratio.   Pane (d) shows the optimized case of $U = \pm 0.2 t$ and bond disorder $W = 1.2 t$, which reaches a conductance ratio of about 120. This value is desirable for constructing  FETs with a clearly observable 'on' / 'off' signal.   If we increase the bond disorder to $W = 1.5 t$ then the 'off' state is driven even deeper into the localized regime and the conductance ratio exceeds 300 in a narrow interval near $E_F = 0.20 t$.   Moreover, in the conductance ratio peaks in pane (d)  the 'on' $G_{+U}$ conductance is in excess of $ 10 e^2/h$ for $U = \pm  0.12t$ and reaches $ 20 e^2/h$  for  $U = \pm 0.20 t$.   This reduces the 'on' conductance's sample to sample variation to acceptable levels.  The  conductance peak reaches from $\sim0.2 t$ to $\sim0.25 t$,  giving it a width of  $\sim0.05 t = 135$ meV. Our conductance ratios are comparable to the  conductance ratios that Xia \textit{et al} ~\cite{xia2010NL} measured by exploiting bilayer graphene's band gap.  They measured a maximum conductance of $26e^{2}/h$,  while their minimum was about $0.26e^{2}/h$ at room temperature and $0.026e^{2}/h$ at 20K.  Our results show that similar conductance ratios can be obtained independently of the band gap  by exploiting disorder and a mobility gap.

The vertical lines in \fref{fig:bilayer} mark the left edge of the peak in the conductance ratio, which occurs when the Fermi level coincides with the bias voltage: $E_F = U$.  Our results for the conductance show that at $E_F = U$ there is a very sharp rise in the 'on' conductance $G_{+U}$; at this Fermi level the effects of disorder are suddenly and strongly reduced.  An opposite effect - a dip - becomes visible  in the 'off' conductance when the  bias voltage is increased to $U = -0.12 t, \, -0.20 t$.  Inside the  conductance ratio peak  the 'off' state moves toward the localized regime, and at the $U = -0.20$  bias it is clearly localized.  These two opposite effects on the 'on' and 'off' conductance both contribute to the peak.  In the next section we will find their origin in the band structure's dependence on the bias voltage.

\begin{figure}
\begin{center}
 \includegraphics[width=\columnwidth,viewport=56 37 702 534, clip]{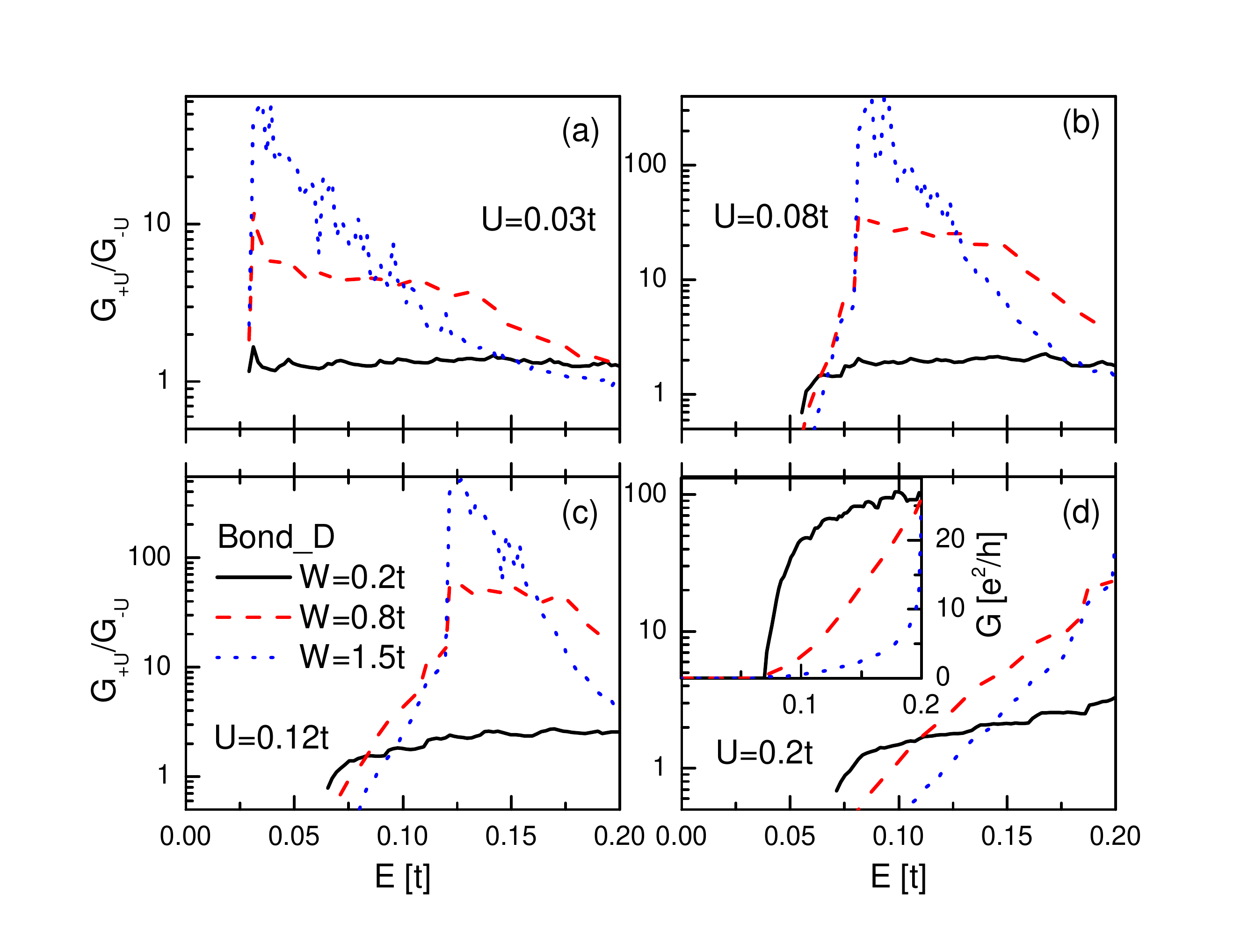}
   \end{center}
   \caption{
   (Color online) Conductance ratio $G_{+U}/G_{-U}$ of bilayer graphene nanoribbons with armchair edges, at three different disorder strengths.   The highest curves were obtained with disorder strength $W = 1.5 t$, while the lowest curves were obtained with $W = 0.2 t$. The four panes report results obtained with four different bias voltages: (a) $U = 0.03 t$, (b) $U = 0.08 t$, (c) $U = 0.12 t$, (d) $U = 0.2 t$.    The $x$ axis shows the Fermi level.
  We do not graph the conductance ratio inside the band gap.  The inset in pane (d) shows the 'on' conductance $G_{+U}$ at $U=0.2t$.}
  \label{fig:armchireBD}
\end{figure}

 \Fref{fig:armchireBD} shows the conductance ratio when the graphene bilayer has armchair not zigzag edges.  The four panes were obtained at four different bias voltages, and within each pane we show three disorder strengths.  Biased armchair-edged bilayer graphene has a band gap, so we omit Fermi levels inside the gap: $|E_F| \leq 0.0279t$ (pane a), $|E_F| \leq0.0547t$ (b), $|E_F | \leq 0.0635t$ (c),  and $|E_F| \leq 0.0701t$ (d).

 \Fref{fig:armchireBD} demonstrates that armchair edges do not cause any fundamental change in the conductance ratio. Again we find that larger conductance ratios are obtained at stronger disorders and larger bias voltages, that the conductance ratio can be optimized to values of order  $O(100)$, and that the left edge of the conductance peak lies at $E_F = U$.   The width of the conductance peak at $W = 0.8 t$, $U=0.08t$ is about $200$ meV.   Changing the edge type  does not significantly change the sample-to-sample reproducibility.  Pane (a) of  \Fref{fig:armchireBD} shows that when the bias voltage is very small $U = 0.03 t$ the conductance ratio exceeds ten only for the very bond large disorder strength $W = 1.5 t$.  Here the bilayer lies near the localized regime both at positive bias $G_{+U}$ and at  negative bias $G_{-U}$, so the 'on' conductance is of order $O(1)$, the same as  its sample-to-sample fluctuations.     This is substantially the same result that we found with zigzag-edged bilayers.   Larger bias voltages $U = 0.08 t, \, 0.12 t, \, 0.20 t$ do produce larger 'on' conductances and reduce the sample-to-sample fluctuations to acceptable levels.  More exactly, we find that when the bond disorder strength is $W = 0.8 t$ the 'on' conductance exceeds ten.  The disorder strength $W = 0.8 t$, like the disorder $W = 1.2 t$ which we examined in zigzag edged bilayers,  puts the  'on' biased system in the conducting regime $G_{+U} > 10$ while forcing the 'off' biased system into the localized regime $G_{-U} \leq 1$.   This is an optimal disorder strength.   The stronger disorder $W = 1.5 t$ shown in  \fref{fig:armchireBD}  localizes the bilayer both in the 'on' and the 'off' state, and therefore reduces the 'on' conductance to small values that do not ensure sample to sample reproducibility.  These numerical results indicate that is necessary to optimize the disorder strength for localization in the 'off' state and conduction in the 'on' state.  Fortunately this optimization requirement is met by a wide range of disorder strengths.

In this section we studied the effect of disorder on the conductance ratio of bilayer graphene under a bias voltage. We found that good FETs with large condutance ratios and small sample-to-sample variation could be obtained when the disorder strength is optimized to moderate values and the bias voltage is not too small~\cite{footnote}. This behavior is independent of  the disorder type and the edge type, but it does depend on the Fermi level. In particular, the left edge of the conductance ratio peak is controlled by the bias voltage; it always appears at $E_F = U$.   In the next section we will show that this behavior originates in graphene's band structure.

\subsection{Band structure of bilayer graphene}

The previous section's results about the conductance ratio can be explained by an analysis of the density of states, which may be either concentrated on one of the two graphene layers or instead shared between them. We will simplify our analysis by using a $\vec{k}\cdot \vec{p}$ continuum model of bilayer graphene.  We use a standard model ~\cite{guinea2006PRB, neto2009RMP} of four bands lying near the  $\overrightarrow{K}$ point:
\[
H_{AB}=\left(
\begin{array}{cccc}
U & \upsilon _{F}k_{-} & 0 & 0 \\
\upsilon _{F}k_{+} & U & t_{\bot } & 0 \\
0 & t_{\bot } & -U & \upsilon _{F}k_{-} \\
0 & 0 & \upsilon _{F}k_{+} & -U%
\end{array}%
\right), \;
P_L=\left(
\begin{array}{cccc}
1 & 0 & 0 &0 \\ 0 & 1 & 0 & 0 \\ 0 & 0 & 0 & 0 \\ 0 & 0 & 0& 0
\end{array}%
\right)
\]
This model's interlayer coupling $ t_{\bot}$ and bias voltage $U$ have the same values as the corresponding parameters in our tight-binding model.  The Fermi velocity is  $\upsilon_{F}=\frac{3a}{2}t$,  and $k_{\pm }=k_{x}\pm ik_{y}$. We will study movement of conduction electrons between the graphene layers using two layer-resolved projection operators: $P_L$ for the lower layer and $P_U = 1 - P_L$ for the upper layer.

\begin{figure}
\begin{center}
 \includegraphics[width=\columnwidth, clip]{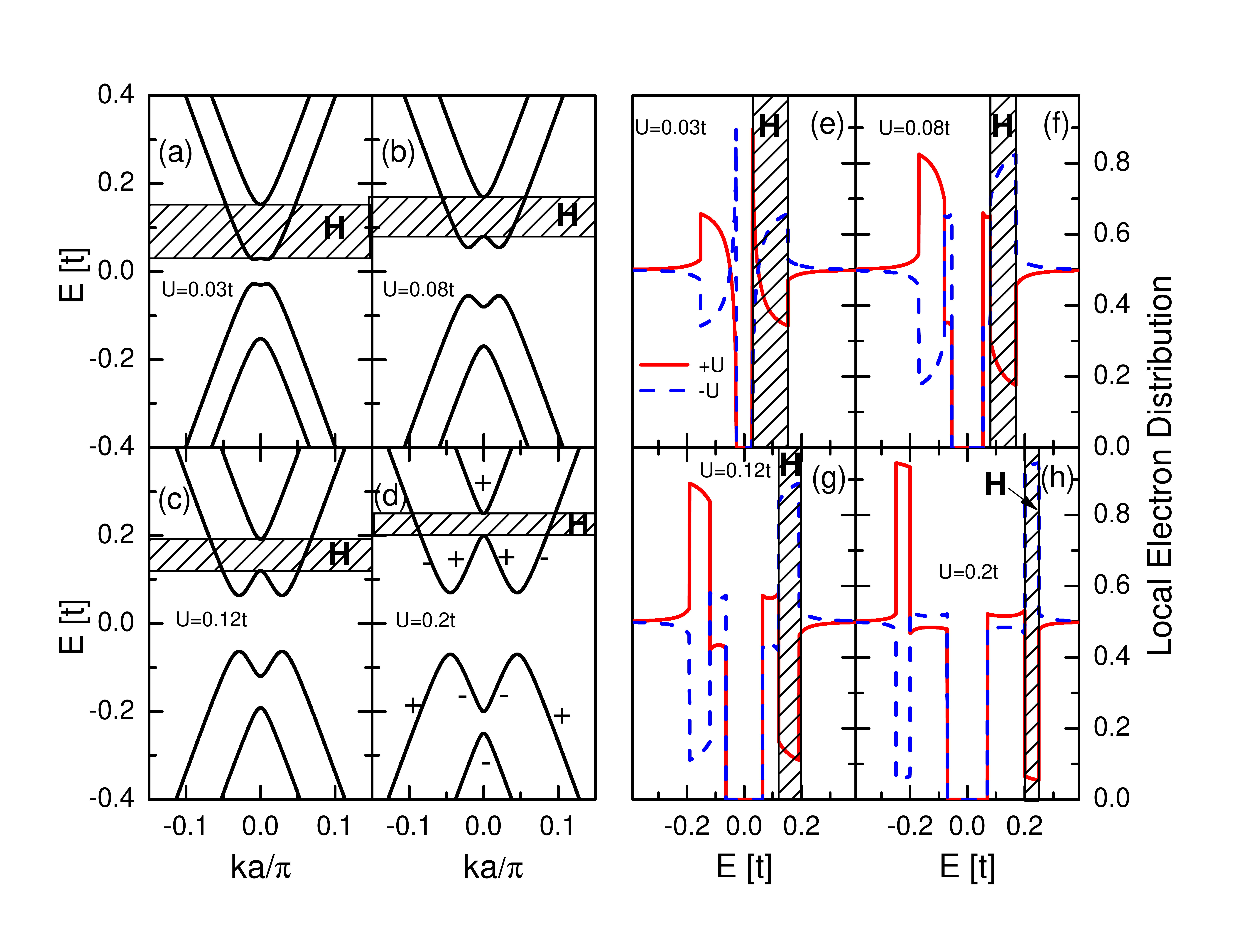}
   \end{center}
   \caption{(Color online.) Band structure of bilayer graphene under a bias voltage $U = 0.03 t, \, 0.08 t, \, 0.12 t, \, 0.20 t$.  The left panes show the energy dispersion, and the right panes show the Local Electron Distribution - the average expectation value of the layer-resolved projection operator.  The plus signs in pane (d) indicate bands associated with the lower graphene layer, and the minus signs indicate the upper graphene layer.   Panes (e)-(h) show the LEDs corresponding to the dispersions in (a)-(d). The red solid lines in panes (e)-(h) plot the lower layer's LED, while the blue dashed lines plot the upper layer's LED. The cross-hatched regions marked with an `H' indicate intervals of the spectrum with only one band that is concentrated on the upper  graphene layer. }
  \label{fig:distributionBilayer}
\end{figure}

In  figure ~\ref{fig:distributionBilayer}  we examine the energy dispersion at four different values of  the bias voltage: $U =  0.03t, \, 0.08t, \, 0.12t,$   and $0.2t$.    When the bias voltage is small (pane (a)) the conduction and valence bands (the second and third bands respectively) are almost flat at small momenta $k_x \approx 0$.  At larger values of the bias voltage (panes (b)-(d)) the conduction band develops a 'Mexican hat'  structure with a peak at $\vec{k} = 0, E_F = +U$ and two minima, and the valence band develops a matching inverted 'Mexican hat' with a valley at $\vec{k} = 0, E_F = -U$ and two maxima ~\cite{guinea2006PRB}.   The minimum gap between the conduction and valence bands, measured from the valleys of the conduction band to the peaks of the valence band, is     $\Delta ^{\prime }=\frac{2Ut_{\bot }}{\left( 4U^{2}+t_{\bot }^{2}\right) ^{1/2}}$  ~\cite{mccann2006PRB}.  When the bias voltage is small $U \ll t_{\bot}$ the gap is proportional to the bias voltage  $\Delta ^{\prime }\approx 2 U $ (as in pane (a)), while when the bias voltage is large  $ U \gg t_{\bot}$ the gap saturates at $\Delta ^{\prime }\approx t_{\bot}$ (as in  pane (d)).

If bilayer graphene's two layers were decoupled ($t_\bot = 0$) then the two bands  of the upper layer would not mix with the two bands of the lower layer.  Each of the four bands would be strictly linear and cross each other at $E_F = 0$.   Instead the two graphene layers do have a nonzero but weak coupling which causes the bands to avoid the $E_F = 0$ crossings and produces the characteristic  'Mexican hat' structure.  Near $E_F =0$ the eigenstates of the bilayer Hamiltonian show significant mixing between the upper and lower graphene layers.   However far from $E_F = 0$ the interlayer coupling and avoided band crossings have little impact, and each of the four bands can be assigned either to the upper graphene layer or to the lower graphene layer.   Figure ~\ref{fig:distributionBilayer}d exhibits this assignment: plus signs indicate the lower layer and minus signs indicate the upper layer.  The uppermost energy band always belongs to the lower graphene layer, but  the conduction band switches allegiances: at small momenta inside the minima of the 'Mexican hat' the conduction band belongs to the lower graphene layer, and at momenta outside the  hat's minima it belongs to the upper graphene layer.

Immediately above the conduction band's  $\vec{k} = 0, E_F = +U$ peak there is a very important gap-like interval separating the conduction band from the uppermost energy band. We have filled this interval with cross-hatched lines and marked it with an H.   Inside it the only electronic states are located in the large-momentum sector of the valence band, and they belong to the upper graphene layer.  States inside this interval concentrate on the upper graphene layer if the bias potential is positive $+U$, and shift to the lower graphene layer if the bias potential is inverted.

The solid red lines in panes (e)-(h) of figure ~\ref{fig:distributionBilayer} plot the Local Electron Density (LED) associated with the lower graphene layer.  This  is the expectation value  $\mathcal{L}_{L,b}(E) =   \langle \psi_b(E) | P_L | \psi _b (E) \rangle $  of the lower layer's projection operator $P_L$, where the index $b$ selects among the bands.  We report\footnote{ Inside the band gap (roughly $E\approx \pm 0.07t$ when $U = 0.2t$) the density of states is zero. For clarity we set the LED's value to zero inside the gap. }  the average over the bands $\sum_b \mathcal{L}_{L,b}(E)/\sum_b$.   If the conduction electrons are concentrated in the lower layer then the LED is close to one, and if they are concentrated in the upper layer then the LED is close to zero.   The dashed blue lines plot the LED $\mathcal{L}_{U}$ associated with the upper graphene layer.  Both lower layer's LED  and the upper layer's LED confirm that inside the cross-hatched intervals a positive bias voltage shifts the conduction electrons from the lower layer to the upper layer.

The LED's sensitivity to a bias voltage explains the peaks which we have seen in the conductance ratio.  When conduction electrons are concentrated in the lower graphene layer they experience a disorder potential which reduces the conductance.  The conductance is further reduced when disorder is increased.   In contrast when conduction electrons are concentrated on  the upper graphene layer they feel a disorder potential that is diminished by the second power of the interlayer hopping $(\frac{t_\bot }{ t})^2 \approx 0.02$, and therefore both the scattering length and the conductance are larger.  In the interval of energies above $E_F = U$  a positive bias voltage shifts conduction electrons to the upper layer and increases the conductance, while a negative bias voltage shifts them to the lower layer and decreases the conductance.    This effect produces the peak in the conductance ratio which we have observed in Figures~\ref{fig:bilayer} and ~\ref{fig:armchireBD}, and explains why its left edge is at $E_F  = U$.  The effect is reversed at bias voltages between the band gap and $E_F = U$: here a positive bias voltage concentrates conduction electrons on the lower layer and  produces the dip in the conduction ratio. Moreover  this effect is absent at large Fermi levels: the LED converges toward  $0.5$ which implies that it is insensitive to bias voltages.  As a result the  conductance ratio is near unity at large Fermi levels.

Further examination of the LED  explains why  the conductance ratio is improved by increasing the bias voltage to large values.   Panes (e)-(h) of figure ~\ref{fig:distributionBilayer}  show that the LED converges toward one (or zero depending on the bias' sign) when the bias is increased; the conduction electrons concentrate more strongly on a single layer, magnifying the disorder's effects.

   It is remarkable that the band structure of pristine bilayer graphene can do such a good job of explaining the conductance ratio peaks and the position of their edges  even though the disorder is quite large.  Thorough explanation of the width and shape of the peaks may require additional consideration, as may also explanation of the conductance ratio at small Fermi levels.  In particular, disorder may renormalize the Fermi level ~\cite{giovannetti2008PRL}, and therefore shift  both the LED and the conductance ratio. Moreover, our conductance calculations were performed on finite size nanoribbons, and the interplay between the localization length scale and the sample size is surely important.

\subsection{Trilayer graphene with two different stacking orders}

\begin{figure}
\begin{center}
 \includegraphics[width=\columnwidth,clip]{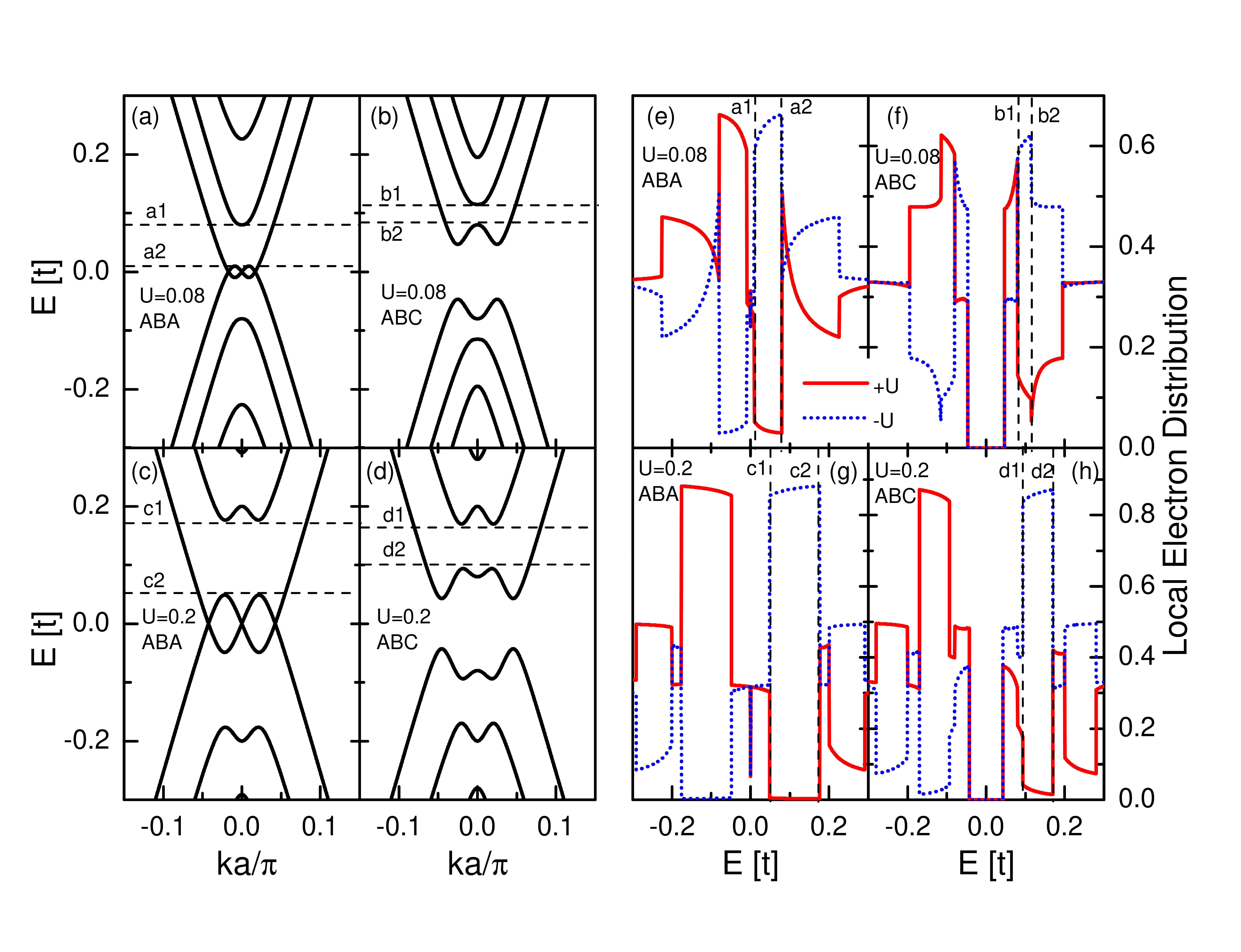}
   \end{center}
   \caption{(Color online) Trilayer graphene's band structure. The left four panes (a)-(d) show the energy dispersion of trilayer graphene under a bias voltage.  The bias voltage is $U = 0.08 t$ for all upper panes and $U = 0.20 t$ for all lower panes.   The left-most panes  (a) and (c) show the ABA (Bernal) stacking's bands.  The adjacent panes (b) and (d)  show the rhombohedral (ABC) stacking's bands and bias-induced band gap.     The right four panes (e)-(h)  plot the Local Electron Distribution - the average expectation value of the layer-resolved projection operator.   The right-most panes were obtained with ABC stacking, and the adjacent panes were obtained with ABA stacking.  The solid red lines give the lower graphene layer's LED, while the dotted blue lines give the upper layer's LED.  A lower level LED near one signals that the conduction electrons are concentrated on the lowest layer.   The black dashed lines (horizontal lines in the left four panes and vertical lines in the right four panes) signal  Fermi levels where a positive bias voltage concentrates the LED on the lowest layer and a negative  bias voltage moves the distribution to the upper layers. }
  \label{fig:distributiontrilayer}
\end{figure}

In the previous sections we have seen that FETs can be fabricated from disordered bilayer graphene.  Trilayer graphene is also an  attractive material.  As mentioned earlier, its  two stable allotropes  show quite different responses to a bias voltage.  The relatively less common rhombohedral (ABC) allotrope exhibits a gap when a perpendicular bias voltage is applied, but the more naturally abundant Bernal (ABA) stacked allotrope does not show any gap.  Research aiming towards FETs and digital logic has focused on the ABC stacking order.  Because the disorder-based FET proposed here does not require a band gap, its application to ABA trilayers is particularly interesting.

We begin by calculating the band structure of both allotropes.  We adopt  $\vec{k} \cdot \vec{p}$ models which are similar to one that we used for bilayer graphene.  They focus on the low energy physics near the $\overrightarrow{K}$ point, and therefore employ only six bands - two for each layer. The $\vec{k} \cdot \vec{p}$ Hamiltonians ~\cite{guinea2006PRB, neto2009RMP}  for trilayer graphene with ABA and ABC stacking are:
\begin{eqnarray}
H_{ABA}=\left(
\begin{array}{cccccc}
U & \upsilon _{F}k_{-} & 0 & 0 & 0 & 0 \\
\upsilon _{F}k_{+} & U & t_{\bot } & 0 & 0 & 0 \\
0 & t_{\bot } & 0 & \upsilon _{F}k_{-} & 0 & t_{\bot } \\
0 & 0 & \upsilon _{F}k_{+} & 0 & 0 & 0 \\
0 & 0 & 0 & 0 & -U & \upsilon _{F}k_{-} \\
0 & 0 & t_{\bot } & 0 & \upsilon _{F}k_{+} & -U%
\end{array}%
\right)
\end{eqnarray}

\begin{eqnarray}
H_{ABC}=\left(
\begin{array}{cccccc}
U & \upsilon _{F}k_{-} & 0 & 0 & 0 & 0 \\
\upsilon _{F}k_{+} & U & t_{\bot } & 0 & 0 & 0 \\
0 & t_{\bot } & 0 & \upsilon _{F}k_{-} & 0 & 0 \\
0 & 0 & \upsilon _{F}k_{+} & 0 & t_{\bot } & 0 \\
0 & 0 & 0 & t_{\bot } & -U & \upsilon _{F}k_{-} \\
0 & 0 & 0 & 0 & \upsilon _{F}k_{+} & -U%
\end{array}%
\right)
\end{eqnarray}
The Fermi velocity and interlayer hopping parameters in these trilayer Hamiltonians are the same as the ones that we employed for bilayer graphene.

Panes (a)-(d) of \fref{fig:distributiontrilayer} display the energy dispersions of these Hamiltonians at two values of the bias voltage.  Our energy dispersions are similar to those obtained in previous studies  ~\cite{kumar2011APL,lui2011NP,bao2011NP}.  When the bias voltage is zeroed both  the ABA stacking and the ABC stacking possess discrete symmetries that guarantee that there is no gap.  The ABA zero-bias dispersion is a combination of linear and quadratic terms, while the ABC dispersion is cubic.  When a bias voltage is imposed the ABC stacking's discrete symmetry is broken, the bands hybridize, and a gap  is clearly visible in panes (b) and (d).    In contrast a bias voltage does not entirely break the ABA stacking's discrete symmetry; the level crossing at the $\overrightarrow{K}$ point is preserved, and panes (a) and (c) show that no gap is opened. The bias does however cause hybridization of the linear and parabolic low-energy bands by breaking the mirror reflection symmetry with respect to the central layer.

Although the spectrum and LED of trilayer graphene are significantly more complicated than those of bilayer graphene, the band structure exhibits similar features that allow a bias voltage to shift the LED between the lower and upper layers. The horizontal dashed lines in panes (a)-(d)  highlight intervals in the spectrum where there is only one band.  When a positive bias voltage is applied,  that band's LED is concentrated outside of the lower layer.  The LED on the lower layer nears zero in the interval between the horizontal dashed lines.    Panes (e)-(h) show  the LEDs of the upper and lower graphene layers. The vertical dashed lines match the horizontal lines in panes (a)-(d), and confirm that in these intervals a positive bias  pushes the lower layer's  LED near to zero and the upper layer's LED near to one.  We conclude that in these intervals the bias shifts conduction electrons between the upper and lower graphene layers, and therefore we expect that  the conductance is sensitive to the bias voltage.

\begin{figure}
\begin{center}
 \includegraphics[width=\columnwidth,viewport=50 32 708 534, clip]{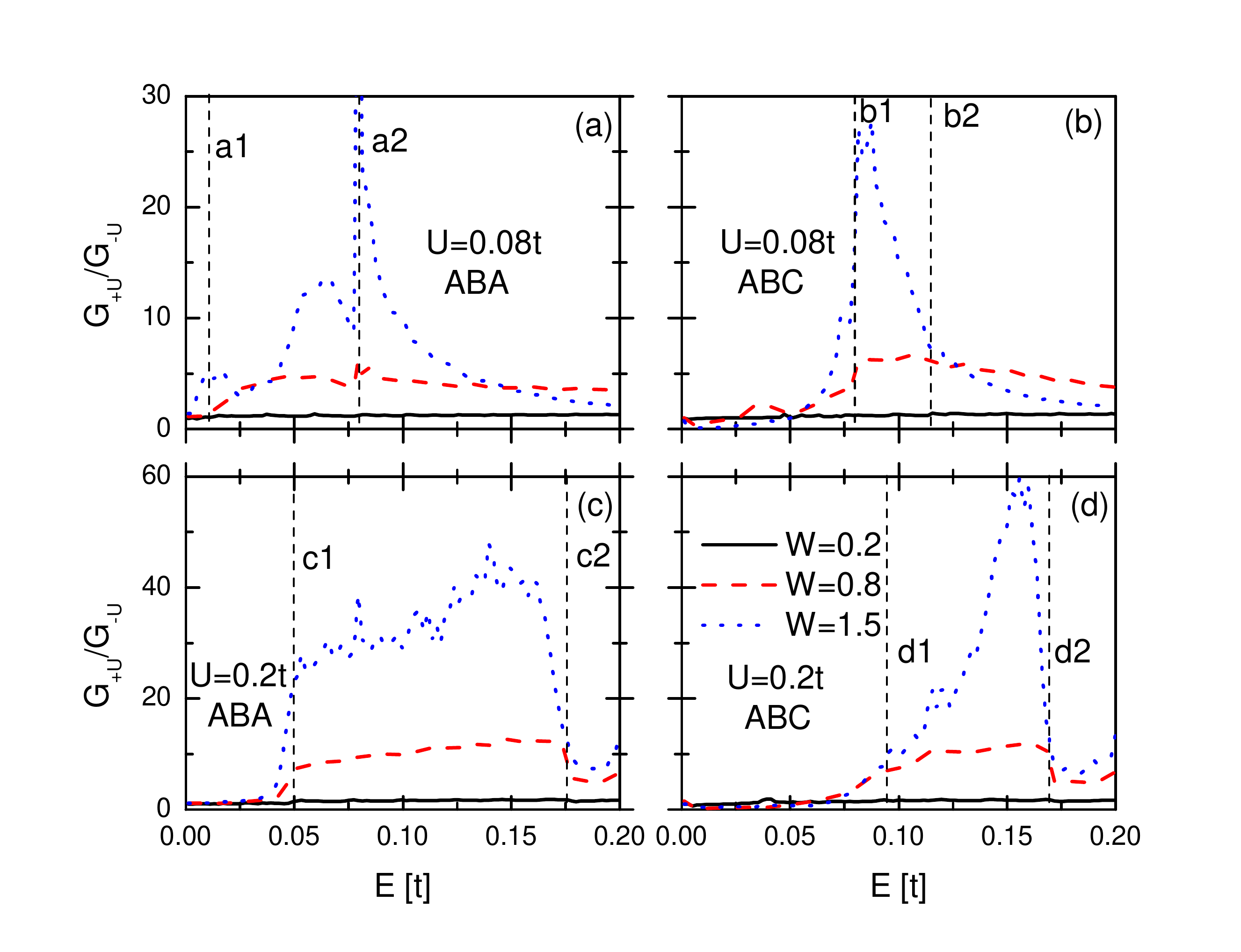}
   \end{center}
   \caption{(Color online.) Conductance ratio of trilayer graphene with ABA stacking (left panes) and ABC stacking (right panes).  The upper panes were obtained with bias voltage $U = \pm 0.08 t$, and the lower panes with $U = \pm 0.20 t$.   Zigzag edges were used in all panes. The  vertical dashed lines are the same as in \fref{fig:distributiontrilayer}, which illustrates the band structure of trilayer graphene.  These lines signal intervals in which a positive bias voltage concentrates the LED on the disordered lowest layer.   }
  \label{fig:tirlayer}
\end{figure}

\Fref{fig:tirlayer} presents the conductance ratio of trilayer graphene with zigzag edges and both stacking orders.  We obtained these results using the tight binding Hamiltonians of equations  \ref{ABATightBinding} and \ref{ABCTightBinding} and the same numerical procedure which we outlined earlier for bilayer graphene.  The dashed vertical lines highlight the intervals where our $\vec{k} \cdot \vec{p}$ Hamiltonians indicate that the LED should be very sensitive  to the sign of bias potential.     As expected, we find  conductance ratio peaks inside these intervals.  The peak width for $W = 1.5 t, U = 0.2 t$ is $\sim0.12 t = 330$ meV in ABA-stacked graphene and  $\sim0.03 t  = 80$ meV in ABC-stacked graphene. We see again that the conductance ratio increases when the disorder strength and/or the bias voltage is increased.  Moreover, even at strong bond disorder $W = 1.5 t$ the 'on' conductance's sample-to-sample reproducibility is good in each pane -  in the conductance ratio peaks $G_{+U}$ ranges  from $5 e^2/h$ to $20 e^2 / h$.  In trilayer graphene the 'on' state is less affected by substrate disorder  than in bilayer graphene, so  the 'on' state conducts even at disorder larger than $W = 1.5 t$.

What impresses us most is that the conductance ratio reaches similar levels in both ABA and ABC stacked trilayers even though no band gap exists in ABA stacked graphene.    Previous research aimed at graphene-based FETs has required a band gap ~\cite{xia2010NL, fiori2009IEEE}.  In this present study of both gapped and ungapped materials  we have established that the conductance ratio does not depend on the presence of a bulk band gap.  Our proposed disorder-based  FET   requires the band structure be sensitive to a bias voltage, so that in the 'off' state the Fermi level coincides with a mobility gap (localized states), while in the  'on' state it coincides with conducting states.    A bias-sensitive mobility gap is all that is necessary.

\section{Conclusion}\label{sec:conlusion}

Our new FET architecture offers considerably more design flexibility and control than  gap-based FETs while still delivering  large conductance ratios and 'off' state widths.  In this present paper we have used simple models to analyze the key physical mechanisms.   Engineering an optimal FET will require more detailed and realistic study concerning the choice of a suitable  substrate, of a preferred disorder type and defect density, and of a method for introducing disorder to the lower graphene layer.  The effects of the substrate and of disorder on the lower layer's structure and chemistry could be better understood, as could the bonding between the graphene layers. Lastly, the influence of the FET's  size and aspect ratio on the conductance ratio deserves some attention.

In summary, we have proposed a new graphene FET architecture which is based on  disorder rather than on a band gap.  We showed that when the bias voltage is large enough and the disorder strength is tuned properly the 'on'/'off' conductance ratio can reach $100$.  The 'on' conductance is large enough to be observed experimentally, and sample-to-sample variations are small enough to ensure reproducible  behavior.  We showed that the FET's sensitivity to a bias voltage is caused by a shift of the conduction electrons from one layer to the other, and that this shifting effect originates in the band structure of pristine multilayer graphene.  We studied bilayer graphene, both natural allotropes of trilayer graphene, and both zigzag and armchair edges, and showed the FET architecture is independent of these details. In particular, our FET architecture does not require a band gap, and requires  disorder concentrated on the bottom layer as an essential element of its design.

 \ack
We thank Xiaoliang Zhong from Michigan Technological University and Dimi Culcer for helpful discussions. This work was supported financially by NBRP of China (2012CB921303 and 2009CB929100), NSF-China under Grants No. 10821403, No.10974236, and No. 11074174, China Post-Doctoral Science Foundation under Grant No. 2012M520099,  and US-DOE under Grant No. DE-FG02-04ER46124.

\section*{References}


\begin{thebibliography}{68}
\bibitem{footnote}
The conductance ratio qualitatively varies in different samples due to conductance fluctuation. However, for majority cases in our simulation, the numerical results show that the conductance ratio is slightly influenced by the sample-to-sample difference, because the conductance fluctuation is much smaller than the on (off) conductance $G_{+U}$ ($G_{-U}$). The physical results in the main text still hold after considering the sample-to-sample difference.
\end{thebibliography}

\end{document}